# CALCULATION OF THE SINGLE-PARTICLE GREEN'S FUNCTION OF INTERACTING FERMIONS IN ARBITRARY DIMENSION VIA FUNCTIONAL BOSONIZATION

PETER KOPIETZ

*Institut für Theoretische Physik, Universität Göttingen*
*Bunsenstr.9, D-37073 Göttingen, Germany*
E-mail: kopietz@physik.uni-goettingen.de

## ABSTRACT

The single-particle Green's function of an interacting Fermi system with dominant forward scattering is calculated by decoupling the interaction by means of a Hubbard-Stratonowich transformation involving a bosonic auxiliary field $\phi^\alpha$. We obtain a higher dimensional generalization of the well-known one-dimensional bosonization result for the Green's function by first calculating the Green's function for a fixed configuration of the $\phi^\alpha$-field and then averaging the resulting expression with respect to the probability distribution $\mathcal{P}\{\phi^\alpha\} \propto \exp[-S_{eff}\{\phi^\alpha\}]$, where $S_{eff}\{\phi^\alpha\}$ is the effective action of the $\phi^\alpha$-field. We emphasize the approximations inherent in the higher-dimensional bosonization approach and clarify its relation with diagrammatic perturbation theory.

## 1. Introduction

The calculation of the single-particle Green's function $G(\mathbf{k}, \omega)$ of an interacting Fermi system in dimensions $d > 1$ is a very difficult problem, which can only be solved approximately. In conventional perturbative many-body theory $G(\mathbf{k}, \omega)$ is calculated by first expanding the irreducible self-energy $\Sigma(\mathbf{k}, \omega)$ to some order in the effective interaction, and then re-summing the perturbation series by solving the Dyson equation [1]. In some cases, however, the perturbation series for $\Sigma(\mathbf{k}, \omega)$ is plagued by divergencies, which can only be cured if infinite orders in the interaction are re-summed. A well known example are interacting electrons in one spatial dimension. Fortunately, in $d = 1$ there exist several non-perturbative methods. Besides the Bethe-Ansatz[2] and renormalization group methods[3], the bosonization approach has been used with great success in one dimension[4]. Anderson's suggestion[5] that the normal-state properties of the high-temperature superconductors are a manifestation of non- Fermi liquid behavior in $d > 1$ have revived the interest to develop non-perturbative methods in higher dimensions. To the best of our knowledge, the first attempt to generalize the bosonization approach to arbitrary $d$ was due to Luther[6]. However, Luther's ideas did not receive much attention until Haldane[7] generalized them in such a way that bosonization in $d > 1$ became a practically useful method for calculating the single-particle Green's function of a large class of interacting Fermi systems. Houghton *et al.* [8] as well as Castro-Neto and Fradkin [9] developed the bosonization approach further and applied it to some problems of physical interest. More recently, we have given an alternative formulation of higher dimensional bosonization[10,11], which is based on a generalization of the functional bosonization approach developed in $d = 1$ by Fogedby[12], and later by Lee and Chen [13]. A similar functional integral approach, which emphasizes more the mathematical aspects of bosonization in $d > 1$, has been developed independently by Fröhlich *et al.* [14].

The advantage of the functional integral approach is that it can be used as a basis to calculate systematic corrections to the non-interacting boson approximation, which exist in any dimension if the energy dispersion is not linearized. Furthermore, in $d > 1$ so called "round-the-corner" processes (this terminology will become evident below) give rise to additional corrections to the leading bosonization result, so that in $d > 1$ bosonization is *certainly not exact*. In this paper we shall describe in detail the derivation of the single-particle Green's function by means of our functional bosonization approach. The approximations inherent in higher dimensional bosonization will be clarified and their validity will be critically discussed. We shall also derive the precise connection between the bosonization result for the Green's function and diagrammatic perturbation theory for the irreducible self-energy. In this way we shall recover a Ward-identity which has first been derived by Castellani, Di Castro, and Metzner[15] by means of a very different method.

## 2. The Hubbard-Stratonowich transformation

The crucial first step in the bosonization approach is the subdivision of the degrees of freedom close to the Fermi surface into sufficiently small boxes $K^\alpha_{\Lambda,\lambda}$ of radial high $\lambda$ and cross section $\Lambda^{d-1}$. The index $\alpha$ labels the boxes in some convenient ordering. The cutoff $\Lambda$ should be chosen sufficiently small so that within a given box the curvature of the Fermi surface can be locally neglected[7-11]. Given this partitioning of the degrees of freedom close to the Fermi surface, a general two-body interaction can be described by the Hamiltonian

$$\hat{H}_{int} = \frac{1}{2V} \sum_{\mathbf{q}} \sum_{\alpha\alpha'} f^{\alpha\alpha'}_{\mathbf{q}} : \hat{\rho}^\alpha_{-\mathbf{q}} \hat{\rho}^{\alpha'}_{\mathbf{q}} : \quad , \tag{1}$$

where $: \ldots :$ denotes normal ordering, and the $f^{\alpha\alpha'}_{\mathbf{q}}$ are coarse grained Landau interaction parameters. Here $\hat{\rho}^\alpha_{\mathbf{q}} = \sum_{\mathbf{k}} \Theta^\alpha(\mathbf{k}) \hat{\psi}^\dagger_{\mathbf{k}} \hat{\psi}_{\mathbf{k}+\mathbf{q}}$ are the Fourier components of the local density operators associated with the boxes, where $\hat{\psi}_{\mathbf{k}}$ is the annihilation operator of an electron with momentum $\mathbf{k}$. The cutoff function $\Theta^\alpha(\mathbf{k})$ is unity for wave-vectors $\mathbf{k} \in K^\alpha_{\Lambda,\lambda}$ and vanishes otherwise. For simplicity, we shall consider spinless fermions. In an Euclidean functional integral approach, the exact Green's function of the system can then be written as[16]

$$G(k) \equiv G(\mathbf{k}, i\tilde{\omega}_n) = -\beta \frac{\int \mathcal{D}\{\psi\} e^{-S_{mat}\{\psi\}} \psi_k \psi^\dagger_k}{\int \mathcal{D}\{\psi\} e^{-S_{mat}\{\psi\}}} \quad , \tag{2}$$

where the Euclidean action $S_{mat}\{\psi\}$ is the following functional of the Grassmann field $\psi$,

$$S_{mat}\{\psi\} = S_0\{\psi\} + S_{int}\{\psi\} \ , \tag{3}$$

$$S_0\{\psi\} = \beta \sum_k [-i\tilde{\omega}_n + \xi_{\mathbf{k}}] \psi_k^\dagger \psi_k \ , \tag{4}$$

$$S_{int}\{\psi\} = \frac{\beta}{2V} \sum_q \sum_{\alpha\alpha'} f_{\mathbf{q}}^{\alpha\alpha'} \rho_{-q}^\alpha \rho_q^{\alpha'} \ . \tag{5}$$

Here $\xi_{\mathbf{k}} = \epsilon_{\mathbf{k}} - \mu$ is the energy dispersion measured relative to the chemical potential $\mu$, and the local density operator is now represented by a composite Grassmann field $\rho_q^\alpha = \sum_k \Theta^\alpha(\mathbf{k}) \psi_k^\dagger \psi_{k+q}$. Throughout this work we shall use the convention that $k = [\mathbf{k}, i\tilde{\omega}_n]$ and $q = [\mathbf{q}, i\omega_m]$, where the fermionic frequencies are $\tilde{\omega}_n = 2\pi(n+\frac{1}{2})/\beta$, and the bosonic ones are $\omega_m = 2\pi m/\beta$. Here $\beta$ is the inverse temperature, and $V$ is the volume of the system.

Defining the dimensionless parameters $\tilde{f}_{\mathbf{q}}^{\alpha\alpha'} = \frac{\beta}{V} f_{\mathbf{q}}^{\alpha\alpha'}$, the interaction can be decoupled by means of the following Hubbard-Stratonowich transformation over a bosonic auxiliary field $\phi_q^\alpha$,

$$\begin{aligned} e^{-S_{int}\{\psi\}} &\equiv \exp\left[-\frac{1}{2} \sum_q \sum_{\alpha\alpha'} [\underline{\tilde{f}}_q]^{\alpha\alpha'} \rho_{-q}^\alpha \rho_q^{\alpha'}\right] \\ &= \frac{\int \mathcal{D}\{\phi^\alpha\} \exp\left[-\frac{1}{2}\sum_q \sum_{\alpha\alpha'} [\underline{\tilde{f}}_q^{-1}]^{\alpha\alpha'} \phi_{-q}^\alpha \phi_q^{\alpha'} - i\sum_q \sum_\alpha \phi_{-q}^\alpha \rho_q^\alpha\right]}{\int \mathcal{D}\{\phi^\alpha\} \exp\left[-\frac{1}{2}\sum_q \sum_{\alpha\alpha'} [\underline{\tilde{f}}_q^{-1}]^{\alpha\alpha'} \phi_{-q}^\alpha \phi_q^{\alpha'}\right]} \ . \end{aligned} \tag{6}$$

Here $\underline{\tilde{f}}_q$ is a matrix in the patch indices, with matrix elements given by $[\underline{\tilde{f}}_q]^{\alpha\alpha'} = \tilde{f}_{\mathbf{q}}^{\alpha\alpha'}$. Eq.(6) is easily proved by shifting the $\phi^\alpha$-field in the enumerator according to $\phi_q^\alpha \to \phi_q^\alpha - i\sum_{\alpha'} [\underline{\tilde{f}}_q]^{\alpha\alpha'} \rho_q^{\alpha'}$ and using the fact that $[\underline{\tilde{f}}_q]^{\alpha\alpha'} = [\underline{\tilde{f}}_{-q}]^{\alpha'\alpha}$, which is a trivial consequence of the hermiticity of the Hamiltonian. With the help of Eq.(6) we obtain from Eq.(2)

$$G(k) = -\beta \frac{\int \mathcal{D}\{\psi\} \mathcal{D}\{\phi^\alpha\} e^{-S\{\psi,\phi^\alpha\}} \psi_k \psi_k^\dagger}{\int \mathcal{D}\{\psi\} \mathcal{D}\{\phi^\alpha\} e^{-S\{\psi,\phi^\alpha\}}} \ , \tag{7}$$

where the decoupled action is given by

$$S\{\psi, \phi^\alpha\} = S_0\{\psi\} + S_1\{\psi, \phi^\alpha\} + S_2\{\phi^\alpha\} \ , \tag{8}$$

with

$$S_1\{\psi, \phi^\alpha\} = \sum_q \sum_\alpha i \rho_q^\alpha \phi_{-q}^\alpha \tag{9}$$

$$S_2\{\phi^\alpha\} = \frac{1}{2} \sum_q \sum_{\alpha\alpha'} [\underline{\tilde{f}}_q^{-1}]^{\alpha\alpha'} \phi_{-q}^\alpha \phi_q^{\alpha'} \ . \tag{10}$$

Thus, the fermionic two-body interaction has disappeared. Instead, we have the problem of a dynamic bosonic field $\phi^\alpha$ that is linearly coupled to the fermionic density. This field mediates the interaction between the fermions in the sense that integration over the $\phi^\alpha$-field (i.e. "undoing" the Hubbard-Stratonowich transformation) generates an effective fermionic two-body interaction. In fact, because all interactions in nature can be viewed as the result of the exchange of some sort of particles, it is more general and fundamental to define the problem of interacting fermions in this way. This point of view has already been emphasized by Feynman and Hibbs[17].

The functional integral approach gives us the freedom of performing the fermionic integration before integrating over the $\phi^\alpha$-field. To eliminate the fermions, we write

$$S_0\{\psi\} + S_1\{\psi, \phi^\alpha\} = -\beta \sum_{kk'} \psi_k^\dagger [\hat{G}^{-1}]_{kk'} \psi_{k'} \quad , \tag{11}$$

where $\hat{G}^{-1}$ is an infinite matrix in momentum and frequency space, with matrix elements given by the formal Dyson equation

$$[\hat{G}^{-1}]_{kk'} = [\hat{G}_0^{-1}]_{kk'} - [\hat{V}]_{kk'} \quad , \tag{12}$$

where $\hat{G}_0$ is the non-interacting Matsubara Green's function matrix,

$$[\hat{G}_0]_{kk'} = \delta_{kk'} G_0(k) \quad , \quad G_0(k) = \frac{1}{i\tilde{\omega}_n - \xi_{\mathbf{k}}} \quad , \tag{13}$$

and the generalized self-energy matrix is

$$[\hat{V}]_{kk'} = \sum_\alpha \Theta^\alpha(\mathbf{k}) V_{k-k'}^\alpha \quad , \quad V_q^\alpha = \frac{i}{\beta} \phi_q^\alpha \quad . \tag{14}$$

Recall that $k$ denotes wave-vector and frequency, so that $\delta_{kk'} = \delta_{\mathbf{kk'}} \delta_{nn'}$. Choosing the normalization of the integration measure $\mathcal{D}\{\psi\}$ suitably, we have then

$$\int \mathcal{D}\{\psi\} e^{-S_0\{\psi\}-S_1\{\psi,\phi^\alpha\}} = det \hat{G}^{-1} = e^{Tr \ln \hat{G}^{-1}} = e^{Tr \ln \hat{G}_0^{-1}} e^{Tr \ln[1-\hat{G}_0 \hat{V}]} \quad , \tag{15}$$

and

$$-\beta \int \mathcal{D}\{\psi\} \psi_k \psi_k^\dagger e^{-S_0\{\psi\}-S_1\{\psi,\phi^\alpha\}} = [\hat{G}]_{kk} e^{Tr \ln \hat{G}_0^{-1}} e^{Tr \ln[1-\hat{G}_0 \hat{V}]} \quad . \tag{16}$$

Hence, after integrating over the fermions the exact interacting Green's function in Eq.(7) can be written as an average of the diagonal element $[\hat{G}]_{kk}$,

$$G(k) = \int \mathcal{D}\{\phi^\alpha\} \mathcal{P}\{\phi^\alpha\} [\hat{G}]_{kk} \equiv \left\langle [\hat{G}]_{kk} \right\rangle_{S_{eff}} \quad . \tag{17}$$

The normalized probability distribution $\mathcal{P}\{\phi^\alpha\}$ is

$$\mathcal{P}\{\phi^\alpha\} = \frac{e^{-S_{eff}\{\phi^\alpha\}}}{\int \mathcal{D}\{\phi^\alpha\} e^{-S_{eff}\{\phi^\alpha\}}} \quad , \tag{18}$$

where the effective action for the $\phi^\alpha$-field contains in addition to the action $S_2\{\phi^\alpha\}$ given in Eq.(10) a contribution due to the fermion determinant,

$$S_{eff}\{\phi^\alpha\} = S_2\{\phi^\alpha\} + S_{kin}\{\phi^\alpha\} \quad , \tag{19}$$

with

$$S_{kin}\{\phi^\alpha\} = -Tr \ln[1 - \hat{G}_0 \hat{V}] \tag{20}$$

Note that in Eq.(17) one first calculates the Green's function for a frozen configuration of the $\phi^\alpha$-field, and then averages the resulting expression over all configurations of this field, with probability distribution given in Eq.(18). The above transformations are exact. Of course, in praxis it is impossible to calculate the exact interacting Green's function from Eq.(17), because (a) the matrix $\hat{G}^{-1}$ cannot be inverted exactly, (b) the kinetic energy contribution $S_{kin}\{\phi^\alpha\}$ to the effective action of the $\phi^\alpha$-field can only be calculated perturbatively, and (c) the probability distribution $\mathcal{P}\{\phi^\alpha\}$ in Eq.(18) is not Gaussian, so that the averaging procedure cannot be carried out exactly. *The amazing fact is now that there exists a physically interesting limit where the difficulties (a), (b) and (c) can all be overcome.* The above method leads then to a new non-perturbative approach to the fermionic many-body problem. The highly non-perturbative character of this approach is evident from the fact that in $d = 1$ the exact solution for the Green's function of the Tomonaga-Luttinger model[4] can be obtained in this way[13,18]. In $d > 1$, this methods leads to a straight-forward generalization of the bosonization approach to arbitrary dimensions. In the next section we shall discuss in detail how this calculation is carried out in praxis.

## 3. Calculation of the Green's function

### 3.1. The Gaussian probability distribution

In general the above kinetic-energy contribution to the effective action can only be calculated perturbatively by expanding

$$S_{kin}\{\phi^\alpha\} \equiv -Tr \ln[1 - \hat{G}_0 \hat{V}] = \sum_{n=1}^{\infty} \frac{1}{n} Tr\left[\hat{G}_0 \hat{V}\right]^n \equiv \sum_{n=1}^{\infty} S_{kin,n}\{\phi^\alpha\} \quad , \tag{21}$$

and truncating the expansion at some finite order. Within *Gaussian approximation* all terms with $n \geq 3$ in Eq.(21) are neglected, so that one approximates

$$S_{kin}\{\phi^\alpha\} \approx Tr\left[\hat{G}_0 \hat{V}\right] + \frac{1}{2} Tr\left[\hat{G}_0 \hat{V}\right]^2 \quad . \tag{22}$$

Carrying out the traces, it is easy to show that within Gaussian approximation

$$S_{eff}\{\phi^\alpha\} \approx i \sum_\alpha \phi_0^\alpha N_0^\alpha + S_{eff,2}\{\phi^\alpha\} \quad , \tag{23}$$

with

$$S_{eff,2}\{\phi^\alpha\} = \frac{V}{2\beta} \sum_q \sum_{\alpha\alpha'} [[\underline{f}_q^{-1}]^{\alpha\alpha'} + \Pi_0^{\alpha\alpha'}(q)]\phi_{-q}^\alpha \phi_q^{\alpha'} \quad , \tag{24}$$

where $\underline{f}_q$ is again a matrix in the patch-indices, with matrix elements given by the Landau parameters defined in Eq.(1). Here $N_0^\alpha = \sum_{\mathbf{k}} \Theta^\alpha(\mathbf{k}) f(\xi_{\mathbf{k}})$ is the number of occupied states in box $K_{\Lambda,\lambda}^\alpha$, where $f(E) = \frac{1}{e^{\beta E}+1}$ is the Fermi function. The polarization part in Eq.(24) is given by

$$\Pi_0^{\alpha\alpha'}(q) = -\frac{1}{V} \sum_{\mathbf{k}} \Theta^\alpha(\mathbf{k}) \Theta^{\alpha'}(\mathbf{k}+\mathbf{q}) \frac{f(\xi_{\mathbf{k+q}}) - f(\xi_{\mathbf{k}})}{\xi_{\mathbf{k+q}} - \xi_{\mathbf{k}} - i\omega_m} \quad . \tag{25}$$

For $|\mathbf{q}| \ll k_F$ we may approximate $\Theta^\alpha(\mathbf{k})\Theta^{\alpha'}(\mathbf{k}+\mathbf{q}) \approx \delta^{\alpha\alpha'}\Theta^\alpha(\mathbf{k})$, so that to leading order in $|\mathbf{q}|/k_F$ we have in any dimension

$$\Pi_0^{\alpha\alpha'}(q) \approx \delta^{\alpha\alpha'}\Pi_0^\alpha(q) \quad , \quad \Pi_0^\alpha(q) = \nu^\alpha \frac{\mathbf{v}^\alpha \cdot \mathbf{q}}{\mathbf{v}^\alpha \cdot \mathbf{q} - i\omega_m} \quad , \tag{26}$$

where

$$\nu^\alpha = \frac{1}{V}\frac{\partial N_0^\alpha}{\partial \mu} = \frac{1}{V}\sum_{\mathbf{k}} \Theta^\alpha(\mathbf{k}) \left[-\frac{\partial f(\xi_{\mathbf{k}})}{\partial \xi_{\mathbf{k}}}\right] \tag{27}$$

is the *local* density of states associated with patch $\alpha$, and $\mathbf{v}^\alpha$ is the local Fermi velocity. Note that the approximation in Eq.(26) is valid for small $|\mathbf{q}|/k_F$ but for *arbitrary* frequencies, and that the local density of states $\nu^\alpha$ is in general a cutoff-dependent quantity. In the work by Houghton *et al.*[8] and Castro Neto and Fradkin[9] it is *implicitly assumed that the Gaussian approximation is justified*. However, in none of these works the corrections to the Gaussian approximation are calculated, so that the small parameter which actually controls the accuracy of the Gaussian approximation remains hidden. Recently we have calculated this parameter by means of our functional bosonization approach[11]. Note that the Gaussian propagator of the $\phi^\alpha$-field is simply given by the RPA-interaction matrix,

$$\left\langle \phi_q^\alpha \phi_{-q}^{\alpha'} \right\rangle_{S_{eff,2}} = \frac{\beta}{V}[\underline{f}_q^{RPA}]^{\alpha\alpha'} = \frac{\beta}{V}\left[[\underline{f}_\mathbf{q}^{-1} + \underline{\Pi}_0(q)]^{-1}\right]^{\alpha\alpha'} \quad , \tag{28}$$

where the elements of the matrix $\underline{\Pi}_0(q)$ are given in Eq.(25).

In the exactly solvable one-dimensional Tomonaga-Luttinger model[4] the bosonized Hamiltonian is known to be quadratic, so that in this case we have *exactly*

$$-Tr \ln[1 - \hat{G}_0 \hat{V}] = Tr\left[\hat{G}_0\hat{V}\right] + \frac{1}{2}Tr\left[\hat{G}_0\hat{V}\right]^2 \quad . \tag{29}$$

All higher order terms vanish identically due to a large scale cancellation between self-energy and vertex corrections, which has been discovered by Dzyaloshinskii and Larkin[19]. A few years later T. Bohr gave a more readable proof of this cancellation[20], and formulated it in form of a theorem which he called the *closed loop theorem*. As shown in Ref.[11], the cancellations responsible for the validity of Eq.(29) in $d = 1$ *exist also in higher dimensions* provided the following two approximations are made[21]:

*(A1): High density-limit or small momentum-transfer limit.*

The interaction should be dominated by momentum transfers $|\mathbf{q}| \lesssim q_c \ll k_F$. Because the fields $\phi_q^\alpha$ mediate the interaction, this condition is equivalent with the statement that the short wave-length Fourier components $\phi_q^\alpha$ with $|\mathbf{q}| \gtrsim q_c$ can be neglected. In this case we may ignore processes that transfer momentum between different patches, provided *the size of the patches is larger than* $q_c$. Note that these "around-the-corner" processes exist for any $d > 1$, because there are always some neighboring patches which can be connected by arbitrarily small momentum transfers. Formally, the existence of the small parameter $\frac{q_c}{k_F}$ justifies the neglect of around-the-corner processes. Evidently, the Gaussian approximation would be completely uncontrolled for the Hubbard model, where fluctuations on all length scales are important.

*(A2): Local linearization of the energy dispersion at the Fermi-surface.*

Suppose we linearize the energy dispersion within a given box $K_{\Lambda,\lambda}^\alpha$ by approximating

$$\xi_{\mathbf{q}}^\alpha \equiv \epsilon_{\mathbf{k}^\alpha+\mathbf{q}} - \mu \approx \mathbf{v}^\alpha \cdot \mathbf{q} \quad , \tag{30}$$

where $\mathbf{k}^\alpha$ is a vector on the Fermi surface that points to the center of patch $\alpha$. Local linearization amounts to replacing Eq.(4) by

$$S_0\{\psi\} \approx \beta \sum_k \sum_\alpha \Theta^\alpha(\mathbf{k})[-i\tilde{\omega}_n + \mathbf{v}^\alpha \cdot (\mathbf{k} - \mathbf{k}^\alpha)]\psi_k^\dagger \psi_k \quad . \tag{31}$$

Note that in this approximation the Fermi surface is approximated by a collection of *flat $d - 1$ dimensional hyper-surfaces*, i.e. planes in $d = 3$ and straight lines in $d = 2$.

If the approximations $(A1)$ and $(A2)$ are valid, then the generalized closed loop theorem discussed in Ref.[11] implies that the probability distribution $\mathcal{P}\{\phi^\alpha\}$ can be approximated by a Gaussian,

$$\mathcal{P}\{\phi^\alpha\} \approx \mathcal{P}_2\{\phi^\alpha\} \equiv \frac{e^{-S_{eff,2}\{\phi^\alpha\}}}{\int \mathcal{D}\{\phi^\alpha\} e^{-S_{eff,2}\{\phi^\alpha\}}} \quad . \tag{32}$$

At zero temperature the first term in Eq.(23) involving the zero wave-vector and frequency component of the $\phi^\alpha$-field can be ignored for the calculation of correlation functions at finite wave-vectors or frequencies.

*3.2. The Green's function for fixed configuration of the $\phi^\alpha$-field*

In order to calculate the Green's function from Eq.(17), we need to invert $\hat{G}^{-1}$. We proceed in two steps. We first show that the condition $(A1)$ means that $\hat{G}^{-1}$ is approximately *block-diagonal*, with diagonal blocks $(\hat{G}^\alpha)^{-1}$ labelled by the patch indices. Therefore the problem of inverting $\hat{G}^{-1}$ can be reduced to the problem of inverting each diagonal block separately.

*Block diagonalization.*

The quadratic form defining the matrix elements $[\hat{G}^{-1}]_{kk'}$ in Eq.(11) can be written as

$$S_0\{\psi\} + S_1\{\psi, \phi^\alpha\} = -\beta \sum_{kq} \psi_k^\dagger [\hat{G}^{-1}]_{k,k+q} \psi_{k+q} \quad , \tag{33}$$

with

$$[\hat{G}^{-1}]_{k,k+q} = \sum_\alpha \Theta^\alpha(\mathbf{k}) \left[ \delta_{q,0}(i\tilde{\omega}_m - \xi_{\mathbf{k}-\mathbf{k}^\alpha}^\alpha) - V_q^\alpha \right] \quad , \tag{34}$$

where $\xi_{\mathbf{q}}^\alpha$ is defined in Eq.(30). The cutoff function $\Theta^\alpha(\mathbf{k})$ groups the matrix elements of the infinite matrix $\hat{G}^{-1}$ into rows labelled by the patch index $\alpha$. To see this more clearly, consider a spherical Fermi surface in $d = 2$. As shown in Fig.1, we partition the degrees of freedom in the vicinity of the Fermi surface into $n$ boxes and label neighboring boxes in increasing order. Evidently the group of matrix elements associated with a given index $\alpha$ in Eq.(34) correspond to the horizontal stripes in the schematic representation of the matrix $\hat{G}^{-1}$ shown in Fig.2(a). The width of the diagonal band with non-zero matrix elements is determined by the range $q_c$ of the interaction in momentum space, because the vanishing of the interaction $f_q^{\alpha\alpha'}$ for $|\mathbf{q}| \gtrsim q_c$ implies that the field $V_q^\alpha$ mediating this interaction must also vanish. Noting that by assumption $(A1)$ $q_c \ll k_F$, we may choose the patch cutoffs $\Lambda$ and $\lambda$ such that

$$q_c \ll \Lambda, \lambda \ll k_F \quad . \tag{35}$$

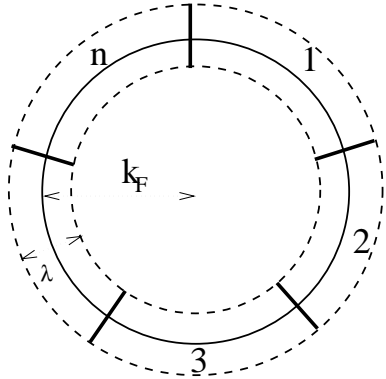

Fig. 1. Subdivision of the degrees of freedom close to spherical Fermi surface in $d = 2$ into boxes.

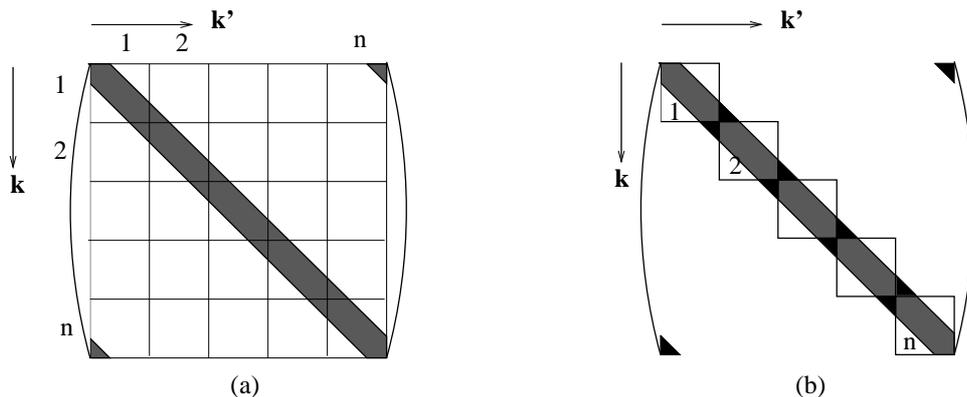

Fig. 2. (a) Schematic representation of the matrix $\hat{G}^{-1}$ defined in Eq.(34) for $d = 2$. Only the wave-vector index is shown, i.e. each matrix-element is an infinite matrix in frequency space. Regions with non-zero matrix elements are shaded. (b) Diagonal blocks and around-the corner processes (represented by black triangles).

As shown in Fig.2(b), in this way the matrix $\hat{G}^{-1}$ is subdivided into block-matrices associated with the patches such that $\hat{G}^{-1}$ is approximately *block-diagonal*. The two black triangles in the upper right and lower left corner of the matrix in Fig.2(b) represent scattering processes between patches 1 and $n$. Because these patches are adjacent, they can be connected by arbitrarily small momentum transfers. The crucial approximation is now to neglect all matrix elements describing momentum transfer between different boxes, i.e. the around-the-corner processes. These are located in the black triangles of Fig.2(b). The justification for this step is that the relative number of matrix elements representing such processes is small as long as the condition (35) is satisfied. In $d > 1$ dimensions the relative number of around-the-corner matrix elements for any given $K_{\Lambda,\lambda}^{\alpha}$ is of the order of $\frac{q_c^d}{\Lambda^{d-1}\lambda} \ll 1$. Note that this approximation makes only sense *if the patch cutoffs are kept finite and large compared with the range of the interaction in momentum space!* Once we have disposed of the matrix elements in the black triangles of Fig.2(b), the matrix $\hat{G}^{-1}$ is a direct sum of diagonal blocks $(\hat{G}^{\alpha})^{-1}$, $\alpha = 1, \ldots, n$, so that

$$[\hat{G}^{-1}]_{kk'} = \sum_{\alpha} \Theta^{\alpha}(\mathbf{k})\Theta^{\alpha}(\mathbf{k}')[(\hat{G}^{\alpha})^{-1}]_{kk'} \quad , \tag{36}$$

where the matrix $(\hat{G}^{\alpha})^{-1}$ is the diagonal block of $\hat{G}^{-1}$ associated with patch $\alpha$,

$$[(\hat{G}^{\alpha})^{-1}]_{kk'} = \delta_{kk'}[i\tilde{\omega}_n - \xi_{\mathbf{k}-\mathbf{k}^{\alpha}}^{\alpha}] - V_{k-k'}^{\alpha} \quad . \tag{37}$$

Thus, *the problem of inverting $\hat{G}^{-1}$ is reduced to the problem of inverting each diagonal block separately*. The diagonal elements of $\hat{G}$ are then simply given by

$$[\hat{G}]_{kk} = \sum_{\alpha} \Theta^{\alpha}(\mathbf{k})[\hat{G}^{\alpha}]_{kk} \quad . \tag{38}$$

Note that $\hat{G}^\alpha$ is still an infinite matrix in frequency space, so that the quantum dynamics is fully taken into account.

Although the relative number of matrix elements describing around-the-corner processes is small, we have to make one important caveat: Possible non-perturbative effects that depend on the *global topology* of the Fermi surface cannot be described within this approximation. For example, in $d = 2$ each patch has two neighbors, but the first and the last patch are adjacent, so that there exist also around-the-corner processes between patch 1 and patch $n$, which give rise to the off-diagonal triangles in the lower left and upper right corners of Fig.2. More generally, in higher dimensions the off-diagonal around-the-corner blocks are distributed in a complicated way over the matrix $\hat{G}^{-1}$. The effect of these sparsely distributed around-the-corner blocks is difficult to estimate, and we are implicitly assuming that they do not lead to qualitatively new effects. We would like to emphasize that this is an *assumption* which is implicitly also made in the operator bosonization approach[8,9], and in the Ward-identity approach by Castellani, Di Casto and Metzner[15].

*Inversion of the diagonal blocks.*

Up to this point we have *not* linearized the energy dispersion, so that the above block diagonalization is valid for arbitrary dispersion $\xi_\mathbf{q}^\alpha$. The crucial advantage of the subdivision of $\hat{G}^{-1}$ into blocks is that within a given block it is allowed to linearize the energy dispersion, $\xi_\mathbf{q}^\alpha \approx \mathbf{v}^\alpha \cdot \mathbf{q}$. This is the approximation ($A2$). By linearizing the energy dispersion, we ignore locally the curvature of the Fermi surface. This is accurate if the patches are sufficiently small, so that the variation of the direction of the local normal vector can be ignored within a given patch. Note, however, that the size of the patches should be chosen large enough to satisfy Eq.(35). Linearization is always justified if the curvature of the Fermi surface is *intrinsically small*, so that even for large $|\mathbf{q}|$ the corrections to $\xi_\mathbf{q}^\alpha \approx \mathbf{v}^\alpha \cdot \mathbf{q}$ can be neglected. In the latter case it is sufficient to cover the Fermi surface with a small number of patches. For example, for quasi-one-dimensional chain-like materials the Fermi surface consists of two disconnected almost flat pieces. As shown in Ref.[22], in this case the magnitude of $\Lambda$ can be chosen of the order of $k_F$.

Once the linearization has been made, it is possible to invert the diagonal block $(\hat{G}^\alpha)^{-1}$ exactly. Note that $\hat{G}^\alpha$ is still an infinite matrix in frequency space. Shifting the wave-vector labels according to $\mathbf{k} = \mathbf{k}^\alpha + \mathbf{q}$ and $\mathbf{k}' = \mathbf{k}^\alpha + \mathbf{q}'$, the diagonal block $\hat{G}^\alpha$ is determined by the equation

$$\sum_{\tilde{q}'} \left[ \delta_{\tilde{q},\tilde{q}'}(G_0^\alpha(\tilde{q}))^{-1} - V_{\tilde{q}-\tilde{q}'}^\alpha \right] [\hat{G}^\alpha]_{\tilde{q}'\tilde{q}''} = \delta_{\tilde{q},\tilde{q}''} \quad . \tag{39}$$

where $G_0^\alpha(\tilde{q}) = [i\tilde{\omega}_n - \mathbf{v}^\alpha \cdot \mathbf{q}]^{-1}$ (For simplicity we have introduced the collective label $\tilde{q} = [\mathbf{q}, i\tilde{\omega}_n]$.) The important point is now that Eq.(39) is first order and can be solved exactly by means of a trivial generalization of a method due to Schwinger[23]. Defining

$$\mathcal{G}^\alpha(\mathbf{r}, \mathbf{r}', \tau, \tau') = \frac{1}{\beta V} \sum_{\tilde{q}\tilde{q}'} e^{i(\mathbf{q}\cdot\mathbf{r} - \tilde{\omega}_n \tau)} e^{-i(\mathbf{q}'\cdot\mathbf{r}' - \tilde{\omega}_{n'} \tau')} [\hat{G}^\alpha]_{qq'} \qquad (40)$$

$$V^\alpha(\mathbf{r}, \tau) = \sum_q e^{i(\mathbf{q}\cdot\mathbf{r} - \omega_m \tau)} V^\alpha_q \quad, \qquad (41)$$

it is easy to see that Eq.(39) is equivalent with

$$[-\partial_\tau + i\mathbf{v}^\alpha \cdot \nabla_\mathbf{r} - V^\alpha(\mathbf{r},\tau)]\mathcal{G}^\alpha(\mathbf{r}, \mathbf{r}', \tau, \tau') = \delta(\mathbf{r} - \mathbf{r}')\delta^*(\tau - \tau') \quad, \qquad (42)$$

where

$$\delta^*(\tau - \tau') = \frac{1}{\beta} \sum_n e^{-i\tilde{\omega}_n(\tau - \tau')} \quad. \qquad (43)$$

Note that the Fourier transformation in Eq.(40) involves fermionic Matsubara frequencies, because $\mathcal{G}^\alpha(\mathbf{r}, \mathbf{r}', \tau, \tau')$ should be anti-periodic in each imaginary time variable. In contrast, $V^\alpha(\mathbf{r}, \tau)$ should be a periodic function of $\tau$, so that $V^\alpha_q$ depends on bosonic Matsubara frequencies. We now substitute the ansatz

$$\mathcal{G}^\alpha(\mathbf{r}, \mathbf{r}', \tau, \tau') = G_0^\alpha(\mathbf{r} - \mathbf{r}', \tau - \tau') e^{\Phi^\alpha(\mathbf{r},\tau) - \Phi^\alpha(\mathbf{r}',\tau')} \qquad (44)$$

into Eq.(42), where $G_0^\alpha(\mathbf{r} - \mathbf{r}', \tau - \tau')$ satisfies

$$[-\partial_\tau + i\mathbf{v}^\alpha \cdot \nabla_\mathbf{r}] G_0^\alpha(\mathbf{r} - \mathbf{r}', \tau - \tau') = \delta(\mathbf{r} - \mathbf{r}')\delta^*(\tau - \tau') \quad. \qquad (45)$$

This yields

$$[-\partial_\tau + i\mathbf{v}^\alpha \cdot \nabla_\mathbf{r} - V^\alpha(\mathbf{r}, \tau)]\mathcal{G}^\alpha(\mathbf{r}, \mathbf{r}', \tau, \tau') =$$
$$\delta(\mathbf{r} - \mathbf{r}')\delta^*(\tau - \tau') + \mathcal{G}^\alpha(\mathbf{r}, \mathbf{r}', \tau, \tau') \{[-\partial_\tau + i\mathbf{v}^\alpha \cdot \nabla_\mathbf{r}]\Phi^\alpha(\mathbf{r}, \tau) - V^\alpha(\mathbf{r}, \tau)\} \quad. \qquad (46)$$

Comparing Eq.(46) with Eq.(42), we see that our ansatz is consistent provided $\Phi^\alpha(\mathbf{r}, \tau)$ satisfies

$$[-\partial_\tau + i\mathbf{v}^\alpha \cdot \nabla_\mathbf{r}]\Phi^\alpha(\mathbf{r}, \tau) = V^\alpha(\mathbf{r}, \tau) \quad. \qquad (47)$$

Eqs.(45) and (47) are first order linear differential equations, which can be easily solved via Fourier transformation

$$G_0^\alpha(\mathbf{r}, \tau) = \frac{1}{\beta V} \sum_{\tilde{q}} \frac{e^{i(\mathbf{q}\cdot\mathbf{r} - \tilde{\omega}_n \tau)}}{i\tilde{\omega}_n - \mathbf{v}^\alpha \cdot \mathbf{q}} \quad, \qquad (48)$$

$$\Phi^\alpha(\mathbf{r}, \tau) = \sum_q \frac{e^{i(\mathbf{q}\cdot\mathbf{r} - \omega_m \tau)}}{i\omega_m - \mathbf{v}^\alpha \cdot \mathbf{q}} V^\alpha_q \quad. \qquad (49)$$

Having determined $G_0^\alpha(\mathbf{r}, \tau)$ and $\Phi^\alpha(\mathbf{r}, \tau)$, the diagonal blocks $(\hat{G}^\alpha)^{-1}$ are inverted, so that $\hat{G}^\alpha$ is explicitly known as functional of the $\phi^\alpha$-field.

*3.3. Gaussian averaging: calculation of the Debye-Waller*

Combining Eqs.(17), (40), (44) and (49), and using the fact that averaging restores translational invariance in space and time, we conclude that the interacting Matsubara Greens-function is given by

$$G(k) = \sum_\alpha \Theta^\alpha(\mathbf{k}) \int d\mathbf{r} \int_0^\beta d\tau e^{-i[(\mathbf{k}-\mathbf{k}^\alpha)\cdot\mathbf{r} - \tilde{\omega}_n \tau]} G_0^\alpha(\mathbf{r}, \tau) \left\langle e^{\Phi^\alpha(\mathbf{r},\tau) - \Phi^\alpha(0,0)} \right\rangle_{S_{eff,2}} \quad . \quad (50)$$

Using Eqs.(14) and (49) we may write

$$\Phi^\alpha(\mathbf{r}, \tau) - \Phi^\alpha(0, 0) = \sum_q \mathcal{J}_{-q}^\alpha(\mathbf{r}, \tau) \phi_q^\alpha \quad , \quad (51)$$

with

$$\mathcal{J}_q^\alpha(\mathbf{r}, \tau) = \frac{i}{\beta} \left[ \frac{1 - e^{-i(\mathbf{q}\cdot\mathbf{r} - \omega_m \tau)}}{i\omega_m - \mathbf{v}^\alpha \cdot \mathbf{q}} \right] \quad . \quad (52)$$

The problem of calculating the interacting Greens-function is now reduced to a multi-dimensional Gaussian integration, which simply yields the usual *Debye-Waller factor*,

$$\begin{aligned}
\left\langle e^{\Phi^\alpha(\mathbf{r},\tau) - \Phi^\alpha(0,0)} \right\rangle_{S_{eff,2}} &= \left\langle e^{\sum_q \mathcal{J}_{-q}^\alpha(\mathbf{r},\tau) \phi_q^\alpha} \right\rangle_{S_{eff,2}} \\
&= \exp\left[ \frac{1}{2} \sum_q \left\langle \phi_q^\alpha \phi_{-q}^\alpha \right\rangle_{S_{eff,2}} \mathcal{J}_{-q}^\alpha(\mathbf{r},\tau) \mathcal{J}_q^\alpha(\mathbf{r},\tau) \right] \\
&= \exp\left[ \frac{\beta}{2V} \sum_q [\underline{f}_q^{RPA}]^{\alpha\alpha} \mathcal{J}_{-q}^\alpha(\mathbf{r},\tau) \mathcal{J}_q^\alpha(\mathbf{r},\tau) \right] \quad , \quad (53)
\end{aligned}$$

where we have used the fact that the Gaussian propagator of the $\phi^\alpha$-field is according to Eq.(28) proportional to the RPA-interaction. For consistency, in Eq.(53) the polarization contribution to $[\underline{f}_q^{RPA}]^{\alpha\alpha}$ should be approximated by its leading long-wavelength limit given in Eq.(26), because in deriving Eq.(53) we have neglected around-the-corner processes. Using

$$\mathcal{J}_{-q}^\alpha(\mathbf{r},\tau) \mathcal{J}_q^\alpha(\mathbf{r},\tau) = \frac{2}{\beta^2} \frac{1 - \cos(\mathbf{q}\cdot\mathbf{r} - \omega_m \tau)}{(i\omega_m - \mathbf{v}^\alpha \cdot \mathbf{q})^2} \quad , \quad (54)$$

we conclude that

$$\left\langle e^{\Phi^\alpha(\mathbf{r},\tau) - \Phi^\alpha(0,0)} \right\rangle_{S_{eff,2}} = e^{Q^\alpha(\mathbf{r},\tau)} \quad , \quad (55)$$

where the *Debye-Waller factor* $Q^\alpha(\mathbf{r}, \tau)$ is given by

$$Q^\alpha(\mathbf{r}, \tau) = R^\alpha - S^\alpha(\mathbf{r}, \tau) \quad , \quad R^\alpha = \lim_{\mathbf{r}, \tau \to 0} S^\alpha(\mathbf{r}, \tau) \quad , \quad (56)$$

with
$$S^\alpha(\mathbf{r},\tau) = \frac{1}{\beta V} \sum_q \frac{f_q^{RPA,\alpha} \cos(\mathbf{q}\cdot\mathbf{r} - \omega_m\tau)}{(i\omega_m - \mathbf{v}^\alpha\cdot\mathbf{q})^2} \quad. \tag{57}$$

Here $f_q^{RPA,\alpha} \equiv [\underline{f}_q^{RPA}]^{\alpha\alpha}$ is the diagonal element of the RPA-interaction matrix. An important special case is a patch-independent bare interaction, i.e. $[\underline{f}_q]^{\alpha\alpha'} = f_\mathbf{q}$ for all $\alpha$ and $\alpha'$. Then it is easy to see that $f_q^{RPA,\alpha}$ can be identified with with the usual RPA-interaction,

$$f_q^{RPA,\alpha} = f_q^{RPA} \equiv \frac{f_\mathbf{q}}{1 + f_\mathbf{q}\Pi_0(q)} \quad, \quad \text{if} \quad [\underline{f}_q]^{\alpha\alpha'} = f_\mathbf{q} \quad, \tag{58}$$

where $\Pi_0(q) = \sum_\alpha \Pi_0^\alpha(q)$ is the total non-interacting polarization.

In summary, we obtain for the full Matsubara Green's function of the interacting many body system
$$G(k) = \sum_\alpha \Theta^\alpha(\mathbf{k})G^\alpha(\mathbf{k} - \mathbf{k}^\alpha, i\tilde{\omega}_n) \quad, \tag{59}$$
where
$$G^\alpha(\tilde{q}) \equiv G^\alpha(\mathbf{q}, i\tilde{\omega}_n) = \int d\mathbf{r} \int_0^\beta d\tau e^{-i(\mathbf{q}\cdot\mathbf{r} - \tilde{\omega}_n\tau)} G^\alpha(\mathbf{r},\tau) \quad, \tag{60}$$
with
$$G^\alpha(\mathbf{r},\tau) = G_0^\alpha(\mathbf{r},\tau)e^{Q^\alpha(\mathbf{r},\tau)} \quad. \tag{61}$$

Shifting in Eq.(59) $\mathbf{k} = \mathbf{k}^{\alpha'} + \mathbf{q}$ and choosing $|\mathbf{q}|$ small compared with the cutoffs $\lambda$ and $\Lambda$ that determine the size of the box $K_{\Lambda,\lambda}^\alpha$, it is easy to see that only the term $\alpha' = \alpha$ in the sum contributes, so that (after renaming again $\alpha' \to \alpha$)

$$G(\mathbf{k}^\alpha + \mathbf{q}, i\tilde{\omega}_n) = G^\alpha(\mathbf{q}, i\tilde{\omega}_n) \quad, \quad |\mathbf{q}| \ll \Lambda, \lambda \quad. \tag{62}$$

## 4. The connection with diagrammatic perturbation theory

In this section we shall derive from Eqs.(59)-(61) an expression for the irreducible self-energy of the many-body system and compare it with the skeleton diagram. In this way we also elucidate the precise relation between higher dimensional bosonization and the Ward-identity approach by Castellani, Di Castro and Metzner (CCM)[15].

*4.1. The integral equation*

Let us apply the differential operator $-\partial_\tau + i\mathbf{v}^\alpha\cdot\nabla_\mathbf{r}$ to the bosonization result for the patch Green's function $G^\alpha(\mathbf{r},\tau)$ in Eq.(61). Using the fact that the application of this operator to $G_0^\alpha(\mathbf{r},\tau)$ generates the usual $\delta$-function (see Eqs.(45) and (43)), it is easy to show that

$$[-\partial_\tau + i\mathbf{v}^\alpha \cdot \nabla_\mathbf{r} + X^\alpha(\mathbf{r}-\mathbf{r}',\tau-\tau')]\,G^\alpha(\mathbf{r}-\mathbf{r}',\tau-\tau') = \delta(\mathbf{r}-\mathbf{r}')\delta^*(\tau-\tau') \quad , \quad (63)$$

with

$$X^\alpha(\mathbf{r}-\mathbf{r}',\tau-\tau') = -[-\partial_\tau + i\mathbf{v}^\alpha \cdot \nabla_\mathbf{r}]Q^\alpha(\mathbf{r}-\mathbf{r}',\tau-\tau') \quad . \quad (64)$$

From the explicit expression for $Q^\alpha(\mathbf{r},\tau)$ given in Eqs.(56) and (57) we find

$$X^\alpha(\mathbf{r},\tau) = \frac{1}{\beta V}\sum_q e^{i(\mathbf{q}\cdot\mathbf{r}-\omega_m\tau)}X_q^\alpha \quad , \quad X_q^\alpha = \frac{f_q^{RPA,\alpha}}{i\omega_m - \mathbf{v}^\alpha\cdot\mathbf{q}} \quad . \quad (65)$$

In Fourier space Eq.(63) becomes

$$[i\tilde{\omega}_n - \mathbf{v}^\alpha\cdot\mathbf{q}]G^\alpha(\tilde{q}) + \frac{1}{\beta V}\sum_{\tilde{q}'} X^\alpha_{\tilde{q}-\tilde{q}'}G^\alpha(\tilde{q}') = 1 \quad , \quad (66)$$

or equivalently

$$[i\tilde{\omega}_n - \mathbf{v}^\alpha\cdot\mathbf{q}]G^\alpha(\mathbf{q},i\tilde{\omega}_n) = 1 - \frac{1}{\beta V}\sum_{\mathbf{q}',n'} \frac{f^{RPA,\alpha}_{\mathbf{q}-\mathbf{q}',i\omega_{n-n'}}}{i\omega_{n-n'} - \mathbf{v}^\alpha\cdot(\mathbf{q}-\mathbf{q}')}G^\alpha(\mathbf{q}',i\tilde{\omega}_{n'}) \quad . \quad (67)$$

Recall that $q = [\mathbf{q}, i\omega_m]$ involves *bosonic* Matsubara frequencies, whereas the label $\tilde{q} = [\mathbf{q}, i\tilde{\omega}_n]$ depends on *fermionic* ones. Because the difference between two fermionic Matsubara frequencies is a bosonic one, the kernel $X^\alpha_{\tilde{q}-\tilde{q}'}$ in Eq.(66) depends on bosonic frequencies. In the zero-temperature limit Eq.(67) is equivalent with the integral equation given in Eq.(13) of the work by Castellani, Di Castro, and Metzner[15]. Our bosonization approach reduces the solution of Eq.(67) to the standard problem of solving a linear partial differential equation (Eq.(46)) and calculating a Debye-Waller factor in a Gaussian integral. In obtaining this solution, the non-trivial atlas with local coordinate systems on the Fermi surface has played an important role. This patching construction is essential to exhibit the large-scale cancellation between self-energy and vertex corrections, which according to the closed loop theorem are guaranteed to happen *in arbitrary d* if the conditions $(A1)$ and $(A2)$ listed above are satisfied. An important difference between our method and the Ward-identity approach of CCM is that these authors do not make use of the patching construction and the associated non-trivial atlas on the Fermi surface. Instead, CCM work with a single rigid coordinate system, with origin at the center of the Fermi sphere. In this coordinate system the integrations seem to be technically more cumbersome. Note also that the final expression for the Green's function derived by CCM explicitly depends on some ultraviolet cutoff, so that it is not obvious how for sufficiently long-ranged interactions the result becomes cutoff-independent. On the other hand, with the patching construction one can take advantage of the fact that for interactions $f_\mathbf{q}$ that are dominated by

|**q**| ≪ $k_F$ only the *local curvature* of the Fermi surface within a given patch determines the accuracy of the approximations. Therefore Eqs.(59)-(61) are accurate *in arbitrary dimension* provided the approximations (A1) and (A2) discussed above are justified. In this case our approach leads to cutoff-independent results for physical correlation functions[10,11,18].

*4.2. The Ward identity*

In diagrammatic perturbation theory it is sometimes convenient[24] to define so called skeleton diagrams in order to exhibit the structure of the perturbation series more clearly. The skeleton diagram for the exact self-energy is shown in Fig.3. In the

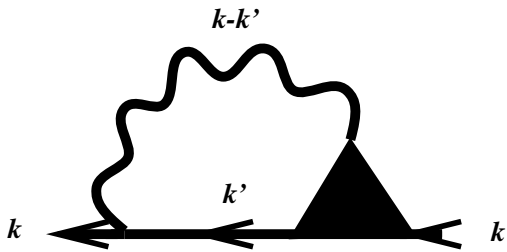

Fig. 3. Skeleton diagram for the irreducible self-energy. The thick wavy line denotes the exact screened effective interaction $f_q^*$, the shaded triangle is the exact three-legged vertex, and the solid line is the exact Green's function.

Matsubara formalism, this diagram represents the following expression,

$$\Sigma(k) = -\frac{1}{\beta V} \sum_{k'} f_{k-k'}^* \Lambda(k; k-k') G(k') \quad . \tag{68}$$

The exact effective interaction $f_q^*$ is related to the bare interaction via $f_q^* = \frac{f_q}{\epsilon(q)}$, where $\epsilon(q)$ is the exact dielectric function. By definition, the vertex function $\Lambda(k;q)$ is the sum of all diagrams with three external ends corresponding to two solid lines and a single interaction line. Because $G(k')$ on the right-hand side of Eq.(68) depends again on $\Sigma(k')$ via the Dyson equation, Eq.(68) is a complicated integral equation, which can only be solved approximately. Moreover, the formal kernel $f_{k-k'}^* \Lambda(k; k-k')$ of this integral equation is again a functional of $G(k)$, so that it cannot be calculated exactly unless the entire perturbation series has been summed.

For better comparison with the self-energy calculated within our bosonization approach, let us shift again **k** = **k**$^\alpha$ + **q** and **k**' = **k**$^\alpha$ + **q**', so that wave-vectors are measured with respect to the local coordinate system $\alpha$. Defining

$$G(\mathbf{k}^\alpha + \mathbf{q}, i\tilde{\omega}_n) = G^\alpha(\tilde{q}) \quad , \tag{69}$$

$$\Sigma(\mathbf{k}^\alpha + \mathbf{q}, i\tilde{\omega}_n) = \Sigma^\alpha(\tilde{q}) \quad , \tag{70}$$

$$\Lambda(\mathbf{k}^\alpha + \mathbf{q}, i\tilde{\omega}_n; \mathbf{q} - \mathbf{q}', i\omega_{n-n'}) = \Lambda^\alpha(\tilde{q}, \tilde{q} - \tilde{q}') \quad , \tag{71}$$

the skeleton equation (68) reads

$$\Sigma^\alpha(\tilde{q}) = -\frac{1}{\beta V} \sum_{\tilde{q}'} f^*_{\tilde{q}-\tilde{q}'} \Lambda^\alpha(\tilde{q}; \tilde{q} - \tilde{q}') G^\alpha(\tilde{q}') \quad , \tag{72}$$

while the Dyson equation can be written as

$$[G^\alpha(\tilde{q})]^{-1} = [G_0^\alpha(\tilde{q})]^{-1} - \Sigma^\alpha(\tilde{q}) \quad . \tag{73}$$

Let us now determine the skeleton parameters that correspond to our bosonization result for the Green's function. Starting point is the integral equation (66). Noting that after linearization $i\tilde{\omega}_n - \mathbf{v}^\alpha \cdot \mathbf{q} = [G_0^\alpha(\tilde{q})]^{-1}$ and dividing both sides of Eq.(66) by $G^\alpha(\tilde{q})$, we obtain

$$[G^\alpha(\tilde{q})]^{-1} = [G_0^\alpha(\tilde{q})]^{-1} + \frac{1}{\beta V} \sum_{\tilde{q}'} \frac{X^\alpha_{\tilde{q}-\tilde{q}'}}{G^\alpha(\tilde{q})} G^\alpha(\tilde{q}') \quad . \tag{74}$$

Comparing this with Eq.(73), we conclude that in our bosonization approach the self-energy satisfies

$$\Sigma^\alpha(\tilde{q}) = -\frac{1}{\beta V} \sum_{\tilde{q}'} \frac{X^\alpha_{\tilde{q}-\tilde{q}'}}{G^\alpha(\tilde{q})} G^\alpha(\tilde{q}') \quad . \tag{75}$$

From Eqs.(72) and (75) we obtain

$$f^*_{\tilde{q}-\tilde{q}'} \Lambda^\alpha(\tilde{q}; \tilde{q} - \tilde{q}') = \frac{X^\alpha_{\tilde{q}-\tilde{q}'}}{G^\alpha(\tilde{q})} = \frac{f^{RPA,\alpha}_{\tilde{q}-\tilde{q}'}}{[i\omega_{n-n'} - \mathbf{v}^\alpha \cdot (\mathbf{q} - \mathbf{q}')] G^\alpha(\tilde{q})} \quad . \tag{76}$$

Hence, the approximations inherent in our bosonization approach amount to replacing the exact effective interaction $f^*_q$ by the RPA-interaction, and approximating the vertex function by

$$\Lambda^\alpha(\tilde{q}; \tilde{q} - \tilde{q}') = \frac{1}{[i\omega_{n-n'} - \mathbf{v}^\alpha \cdot (\mathbf{q} - \mathbf{q}')] G^\alpha(\tilde{q})} \quad , \tag{77}$$

which is equivalent to

$$[i\omega_{m'} - \mathbf{v}^\alpha \cdot \mathbf{q}'] \Lambda^\alpha(\tilde{q}; q') = [G^\alpha(\tilde{q})]^{-1} \quad . \tag{78}$$

The important point is that the left hand side of Eq.(78) depends again on the exact Green's function. Such a relation between a vertex function and a Green's function is called a *Ward identity*. Using the definition (69) and the fact that after linearization $G_0^\alpha(\tilde{q}) = -G_0^\alpha(-\tilde{q})$, it is easy to see that Eq.(78) is equivalent with the Ward identity derived by CCM[15]. Thus, although within the bosonization approach the dielectric function is approximated by the RPA-expression, bosonization does not simply reproduce the usual RPA self-energy, because it sums in addition infinitely many other diagrams by means of a non-trivial Ward-identity for the vertex function! The analytic expressions for these diagrams can be easily obtained by iterating the integral equation (67). Let us recapitulate how we have obtained Eq.(78): Starting point was the bosonization result for the Green's function of the interacting many body system in Eq.(61). By simple differentiation we have obtained the integral equation (67). Finally, we have shown that this integral equation implies the Ward identity (78). The strategy of CCM was to perform these steps in precisely the opposite order.

## 5. Conclusions

In this work we have shown that by simple Gaussian integration one can obtain a non-perturbative expression for the single-particle Green's function which sums infinite orders in perturbation theory in a consistent way. Our approach is controlled in the high density limit for arbitrary dimensions if the interaction is dominated by small momentum transfers. The fact that in one dimension we obtain the exact solution of the Tomonaga-Luttinger model shows that our approach does not assume a priori that the system is a Fermi liquid. Thus, with the method described in this paper it is possible to study strongly correlated Fermi liquids with small quasi-particle residue, as well as possible non Fermi liquid states in higher dimensions. For example, we have applied our functional bosonization approach to quasi-one-dimensional metals[22], to electrons coupled to transverse gauge fields[25], to strongly coupled electron-phonon systems[18], and to the problem of electrons moving in a stochastic medium[18]. Although the derivation of Eqs.(59)-(61) was rather straight-forward, the evaluation of the full momentum- and frequency dependent spectral function from these expressions is a very difficult mathematical problem, which so far has not been completely solved. A detailed discussion of Eqs.(59)-(61) and applications to problems of physical interest will be published in a forthcoming book[18].

## 6. Acknowledgements

The functional bosonization approach described in this work was developed in spring 1994 in collaboration with Kurt Schönhammer. I would also like to thank him for his comments on this manuscript. I am grateful to the organizers of the Raymond L. Orbach symposium for giving me the opportunity of presenting this approach at the symposium, and the Deutsche Forschungsgemeinschaft for supporting my participation at the symposium with a travel grant. I have enjoyed very much my visit to Riverside, especially the great dinner reception at Ray's residence.